\documentstyle[preprint,prl,aps]{revtex}
\begin{document}

\title{A Readout System for the STAR Time Projection Chamber}

\author{M. Anderson$^1$, R. Bossingham$^2$, F. Bieser$^2$,
D. Cebra$^2$, E. L. Hjort$^3$, \\ S. R. Klein$^2$,  C. Q. Vu$^2$,
H. Wieman$^2$}

\address{
$^1$University of California, Davis, CA, 95616, USA\break
$^2$Lawrence Berkeley National Laboratory, 
Berkeley, CA 94720, USA\break 
$^3$Purdue University, West Lafayette, IN, 47907, USA\break}

\break 
\maketitle
\vskip -.2 in
\begin{abstract}
\vskip -.2 in 

We describe the readout electronics for the STAR Time Projection
Chamber.  The system is made up of 136,608 channels of waveform
digitizer, each sampling 512 time samples at 6-12 Mega-samples per
second.  The noise level is about 1000 electrons, and the dynamic
range is 800:1, allowing for good energy loss ($dE/dx$) measurement
for particles with energy losses up to 40 times minimum ionizing.  The
system is functioning well, with more than 99\% of the channels
working within specifications.

\end{abstract}
\narrowtext

\section{Introduction}

The Solenoidal Tracker at RHIC (STAR) is a large detector, designed to
study relativistic heavy ion collisions at center of mass energies up
to $\sqrt{S_{NN}} = 200\ $GeV per nucleon, for nuclei ranging from
protons to gold\cite{STAR}. The main purpose of the experiment is to
search for signals of the quark-gluon plasma (QGP).  A wide variety of
QGP signatures have been proposed.  STAR is optimized to study
produced hadrons over a large fraction of phase space.  This imposes
many requirements on the central time projection chamber (TPC) that is
the central detector for STAR and on its front end electronics (FEE).
The TPC must track and identify particles over as large an acceptance
as possible.  Because central heavy ion collisions produce
high-multiplicity final states, it must have good pattern recognition
and reconstruction capabilities in regions that are densely packed
with tracks.  The tracking must be efficient at finding and
identifying low-momentum particles, and have good momentum resolution
for high-momentum ones.  It must be able to accurately point tracks
back to the primary vertex and be able to detect secondary and
tertiary vertices such as $K^0$, $\Lambda$, $\Xi$ and $\Omega$ decays.
It must be able to identify particles by $dE/dx$ below the
relativistic rise.  In order to measure the source size via
Hanbury-Brown Twiss correlations, it must have excellent two-track
separation.  In addition to the search for the QGP, RHIC will also
study collisions of polarized protons and also ultra-peripheral
electromagnetic interactions involving heavy ions.

The STAR TPC is a 4 meter long cylinder surrounding the beam
pipe\cite{TPC}. The TPC covers from 50 cm to 200 cm in radius.  The
gas is 90\% argon, 10\% methane (P-10) at 1 atmosphere; 50\% helium,
50\% ethane is a possible upgrade to reduce multiple scattering. The
electronics is designed to accommodate either gas.

Electrons from ionized gas drift toward the nearest endcap, where they
are read out.  The TPC readout is divided into 24 sectors, 12 at
each end.  Each sector is divided into inner and outer subsectors.  As
Fig.  \ref{fTPCRD} shows, the readout chambers have 3 wire planes: a
gated grid, ground plane, and anode wires.  The inner and outer sector
geometries are somewhat different, to emphasize 2-track separation at
small radii, and good $dE/dx$ and moderate channel count at large radii.
Table I compares the inner and outer sector characteristics.  

The gated grid is normally closed, minimizing buildup of positively
charged ions in the drift volume.  When a trigger is received, the
voltages are switched and it becomes transparent.  The ground plane,
anode wires and TPC pad plane form a multi-wire proportional chamber.
Electrons drift to the wire where they initiate an avalanche, leaving
a cloud of positively charged ions remaining around the wire.  The
136,608 pads on the TPC pad plane image this charge; this image charge
goes to the electronics.

The size and shape of the ion cloud depends on the number of primary
ions, drift distance and diffusion, and gas gain.  These factors are
quite different for the inner and outer sectors.  The signal $S$
reaching the electronics depends on the capacitative coupling $C_c$
between the ion cloud and the pad:
\begin{equation}
S = N_{pri}GC_cT
\end{equation}
where $N_{pri}$ is the geometry-dependent number of primary electrons,
$G$ is the gas gain, and $T=0.63$ accounts for the convolution of the
diffusion-spreading with the preamplifier shaping.  The value 0.63 is
the convolution of a Gaussian distribution for a 2 meter drift
distance with the actual amplifier response function.  Table 1
compares the other values for the inner and outer sectors.

The gas gains depend on the wire voltages $V$.  These gas gains were
measured in chamber prototypes using an $^{55}Fe$ source.  The inner
($G_i$) and outer ($G_o$) sector gains may be parameterized
\begin{eqnarray}
G_i = \exp{[0.01267(V-520.1)]} &\ \ \ \ \ \ \ \ & 
G_o = \exp{[0.009341(V-628.7)]}.
\end{eqnarray} 
Since the signal has a long tail as the ions drift away from the wire,
the measured gain depends on the shaping time used in the amplifier.
The quoted gains are for the 200 ns FWHM shaper used in STAR. 

To minimize the charge deposited on the wires, and maximize the
chamber lifetime, the anode wires are operated at the lowest possible
gain consistent with a 20:1 signal to noise ratio: about 3770 in the
inner sectors, and only 1230 in the outer sectors.  Since the noise
levels are very similar, the two geometries produce similarly sized
minimum ionizing particle signals, about 22,000 electrons.

The width of the signals depends on both the diffusion and on the
response of the wire chamber.  In P-10 gas in the full 0.5 T magnetic
field, the transverse and longitudinal diffusion are 185 $\mu
m/\sqrt{cm}$ and 320 $\mu m/\sqrt{cm}$ respectively.  After a typical
1 meter drift, an electron cloud measures 1.8 mm transversely by 3.2
mm longitudinally.  This cloud is further broadened by the track dip
angle and crossing angles.  The TPC resolution is roughly given by the
diffusion divided by $\sqrt{N_{pri}}$ where $N_{pri}$ is the number of
primary electrons, typically 100.  After accounting for the dip and
crossing angles and averaging over chamber position, the resolution is
about 350 $\mu$m transversely by 700 $\mu$m longitudinally.  The TPC
TPC pads have widths far larger than this transverse resolution.
Because of this, a multi-pad fit is required to reach the diffusion
limit.

The STAR system\cite{oldfee} is similar to the design used for the
EOS\cite{EOS} and NA-49\cite{NA49} TPC electronics.  In fact, NA-49
and STAR share the SCA/ADC chip design.  However, the STAR system has
a larger dynamic range, lower noise, and a much faster readout rate.
A slightly modified version of this electronics is also used for the
22,000 pads in the 2 STAR forward TPCs (FTPCs)\cite{FTPC}.

\section{Electronics Requirements \& Specifications}

Because of the high probability of multiple hits on a single TPC pad,
each pad must be read out with a waveform digitizer.  If the system
takes several samples for each ionization cluster reaching the pad,
then a 3-point fit can be used to localize the hit to within a
fraction of a time sample.  For a 3-point fit, calculations show that
the contribution of the electronics noise to the resolution is
comparable to the intrinsic chamber resolution when the signal to
noise ratio is 20:1.  This requires that the electronics noise must be
less than 1,000 electrons.  Noise decreases with increasing shaping
time; the electronics is best matched to the TPC if the shaping time
is comparable to the width of the signals.  With P-10, the drift
velocity is about 5.5 cm/$\mu$s (36$\mu$s total), so signals are about
200 ns FWHM.  Helium-ethane has a much slower, 2.5 cm/$\mu$s drift
velocity (80$\mu$s total), but less diffusion, so the change in signal
length is moderate. To accommodate the variation, a shaper with a
adjustable shaping time was chosen.  It should be noted that the
chamber resolution could be improved slightly by increasing the signal
to noise ratio further.

The other significant analog challenge came from the wire response
function.  The signals have a long-lived ($\sim 20 \mu$s) tail due to
the time it takes for the positive ion cloud to disperse from around
the wire.  This tail is different for P-10 and helium-ethane.  This
tail is removed with an electrically adjustable tail cancellation
circuit.  The response was fine tuned by using digitized waveforms of
the actual chamber wire response function.

The number of digitized samples is set by the diffusion spreading and
two-track resolution.  If the signal is spread over 3 buckets, then
the arrival time can be found to $\sim 20\%$ of a bucket; this
requires about 500 time samples.  The electronics accepts a maximum of
512 samples (with an option to upgrade to 1024).  In the summer 2000
run about 380 time-buckets were filled.  The multi-pad and multi-time
sample fits used to find the position of the ionization clusters
impose fairly stringent requirements on channel to channel gain
variation ($<5$\%) and crosstalk ($<0.3$\%).

The dynamic range is set by the energy loss ($dE/dx$) of produced
particles. The maximum ionization expected was from a 200 MeV/c
proton, which produces a signal that is 10 times minimum ionizing.  We
allow a further factor of 4 to account for non-zero dip angles (which
increase the path length) and Landau fluctuations.  With the 20:1
signal to noise ratio, this leads to an 800:1 dynamic range, or 10
bits.

Finally, the system must work in a 0.5 T magnetic field.  Cooling
considerations and the need to maintain the pad plane at a constant
temperature (for $dE/dx$ stability) limit the power consumption to
less than 100 mW/channel.  The system must be read out in 10 msec.
Although STAR will write less than 100 events/second to tape, the
faster readout rate will allow a larger event sample to be examined by
the level 3 trigger.

The FTPC FEE requirements were similar to those for the main TPC.
However, the signals were considerably slower, and so the signal
shaping required a FWHM of about 400 ns.  The longer shaping time
reduced both the signal slewing rates and the preamplifier noise. Only
256 time-buckets were needed in the system.  However, because of the
limited cooling available, power consumption was a problem.  Because
of the slower rise times, we were able to run the SAS preamplifiers
and output buffers at considerably lower currents than in the main
TPC, 40 mW/channel.

\section{System Architecture}

Figure \ref{foverview} shows how the system is divided into two
components.  Small, 32-channel FEE cards plug into the TPC pad plane,
and contain all of the circuitry for amplification, shaping and analog
to digital conversion.  The FEE cards are supported by larger readout
boards, which provide power and control signals, read out the data,
and send it to the data acquisition system.

The FEE card contains two custom chips: a STAR preamplifier/shaper
(SAS) chip and a switched capacitor array/ADC chip (SCA/ADC); each
chip contains 16 channels.  The SAS contains low noise integrating
preamplifiers, 2-pole shapers and buffers to drive the SCA chip.  The
SCA/ADC includes 512 time-bin switched capacitor arrays and 12-bit
Wilkinson ADC's, plus an output buffer and multiplexer. Both chips are
implemented in a 1.2$\mu$m 2-poly, 2-metal CMOS process.  A key
attribute of this process was the ability to make high-quality (good
dielectric) capacitors for the SCA.

Up to 36 FEE cards are controlled and read out by a readout board
(RDO).  The board multiplexes the digital signals from the FEE cards
and sends the data to the data acquisition system (DAQ) on a 1.2
Gbit/sec fiber optical link.  A trigger system initiates events,
including several types of calibrations, and distributes the SCA
acquisition clocks.  The readout board also includes a 1-event memory
for event storage (mostly for debugging), a slow controls link that
can turn parts of the system on and off and monitor temperatures and
voltages.  It also includes voltage regulators for itself and the FEE
cards.

In the FTPCs, the FEE cards each read out 64 channels; a single FEE
card is connected to a cable.  By using smaller chip packages, the
FTPC FEE cards are about the same size as their main TPC counterparts,
despite the increased channel count.

\subsection{SAS chip}

Fig. \ref{fSAS} shows a block diagram of the SAS chip\cite{Ericb}.
The input is a folded cascode integrator with a switched reset.  The
input FET capacitance is about 10 pF, optimized to match the TPC pad
capacitance.  The FET measures 7200 $\mu$m by 1.2$\mu$m, and runs with
2.9 mA drain current.  This gives it a noise of 560 electrons + 14
electrons per pF of input capacitance.  The typical input capacitance
was about 25 pF, from the pad, connecting trace and connector.  This
led to a typical noise level of 910 electrons.

The integration capacitor $C_i$ is 1.6 pF, big enough to handle a
maximum input charge of 1.1 pC, or 300 minimum ionizing particles.  A
resistive discharge was considered for the integrator, but was
abandoned because of the difficulty of obtaining consistent,
appropriately high resistance.  When a trigger occurs, the reset
switch opens in parallel with the TPC gating grid, so no data is lost
due to the opening time.

The chip includes input protection diodes for protection against
sparking.  Although the FEE cards include a place for external
protection diodes for additional protection, external diodes were not
installed.  If an anode wire sparks to a pad, damage will usually be
limited to that channel, with some chance of destroying the entire
16-channel SAS chip.  This was considered an acceptable risk, compared
to the problems of leakage currents in the external diodes.

The two-pole shaper and tail correction are both electrically
adjustable to accommodate either P-10 or helium-ethane gas.  The shaper
peaking time is adjustable from 60 to 150 ns.  The tail correction
timing is similarly adjustable.  The tail correction is adjustable so
that, for either P-10 or He-Eth, after $1\mu$s, the remaining
uncorrected tail is less than 1\% of the maximum input.

The resistors are implemented with MOSFETs biased in their linear
region.  The resistances are adjustable by changing the gate voltage;
these voltages are externally adjustable.  The drawback of this
approach is that the resistors are only linear over a limited range,
usually $<1$ volt.  So, the resistors were implemented with 4 MOSFETs
in series.  With this circuit, the width changes by $<5\%$ going from
small signals to full scale.  The chip-to-chip variation in timing
constants and chip gain are both less than 4\%.  Unfortunately, the
chip-to-chip gain variation is largely correlated with the variation
in pulse width, so that the integrated output voltage varies by a
larger factor, up to 8\%.

The shaper amplifier uses the `cold resistor` technique for
differentiation.  This reduces the noise referred to the input. In the
SAS, almost all of the noise comes from the input FET.

The overall chip gain is 16 mV/fC, so a minimum ionizing signal is
about 40 mV; the maximum output voltage is 2 volts.  The crosstalk
specification is $<$ 0.36\%.  This required using very heavy
metallization. 

The chip includes a charge-injection calibration system.  Charge can
be injected into each input through a capacitor connected to an
external voltage.  A serially loaded shift register determines which
channel(s) are pulsed.

The output buffer is designed to drive a 50 pF load (2 SCA's). The
reason for this is that the FEE card includes an option to add 2
additional SCA/ADC chips to the back of the board, to expand the
system to 1024 time samples.

The SAS die measures 2.6 mm by 3.6 mm and was packaged in an 68 pin
PLCC.  The yield was about 70\%, in accord with expectations based on
the die size.  Dice were also mounted in an TQFP80 package for use in
the FTPC readout.

\subsection{SCA/ADC chip}

As Fig. \ref{fSCA} shows, each channel of the SCA/ADC chip is made up
of a 512-capacitor array which serves as an analog storage
unit\cite{SCA}, and a 12 bit ADC\cite{ADC}.  During sampling, each
capacitor is connected to the chip input in turn, by a switch.  The
512 switches are closed in turn by a shift register.

The SCA performance is dependent on the process used to make the
chip. We used a 1.2 $\mu$m CMOS process; the 0.7 pF capacitors are made from
two metal layers separated by a polysilicon dielectric.  This gives
them a low dielectric loss angle. 

The ADC is a conventional 12-bit Wilkinson rundown ADC; STAR uses only
10 bits.  In STAR, it counts on both edges of a 90 MHZ clock signal.
Counting takes about 6$\mu$s, with an additional $\sim 4$ $\mu$s
allowed for the op amp settling and switching between capacitors.
Running 16 converters in parallel introduces significant economies of
scale. Each individual channel requires only a comparator and a latch.
A single voltage ramp feeds all 16 comparators and a single 12-bit
counter feeds all 16 latches.  The 12-bit counter is implemented as 3
4-bit Grey code counters.  This design was a tradeoff between circuit
complexity and minimizing the number of transitions on any given clock
edge.  During SCA testing, the chips are tested with a slow ramp.  Any
problems in the Grey code counter will show up as non-monotonic
behavior on the ramp, and the chip will be rejected.  The 16 latches
are read out by a double-buffer, a 16:1 multiplexer, a Grey code to
binary converter, and an output buffer.  The buffer was designed to
allow analog to digital conversion to proceed during readout of the
preceding time samples.  We do not use this feature because it
introduces considerable additional noise.

The system has a dynamic range of up to 4 volts; we use a 2 volt range
because it offers improved linearity, $<2$\%.  One ADC count
corresponds to 2 mV.  The SCA noise is about 1 mV, relatively
independent of operating condition.  This noise is probably due to
charge injection in the switches.  The pedestal varies by about 4 mV
from time-bin to time-bin; this pedestal is subtracted on-line by the
data acquisition system.

The analog performance of the SCA/ADC was studied with a variety of
methods.  The dynamic linearity was studied with fast (5 MHz) triangle
waves, and was found to match the static behavior.  The effective
aperture time was studied with an input step function.  The phase of
the step with respect to the sample clock was adjusted in 1 ns
steps, and the changing output monitored.  The measured rise time
was about 5 ns, comparable to the rise time of the input step.  This
was more than satisfactory for STAR needs.  Crosstalk was observed
only between adjacent channels, but most of this is probably external
to the chip.

The biggest chip-to-chip variation was in the SCA/ADC gain, about $\pm
15\%$.  The channel-to-channel gain variation within a chip was small,
typically less than 0.4\%.  The larger chip-to-chip gain variation was
likely in the ADC ramp generator. The chip-to-chip gain exceeded our
specifications, so we sorted the chips by gain, into 6 different
classes.  A different external resistor was used in the ramp generator
for each class, to greatly reduce the gain variation.

Because of the DAQ zero suppresion design, all pedestals were required
to be less than 256 ADC counts (512 mV)\cite{DAQ}.  Most pedestals
were in the range of 125-175 ADC counts.  Within a channel, each
time-bin had a slightly different pedestal; typical variation was a
few mV.  These pedestals were determined and subtracted by the data
acquisition system.  The pedestals have been very constant with time,
varying by a small fraction of an ADC count.

If the system is dormant for too long (10's of seconds), leakage
allows the voltages on the SCA capacitors to drift, eventually
reaching saturation. When they reach saturation, a single readout
cycle may not be enough to fix the capacitors to the correct input
voltage.  Because of this, we do not accept the first event after the
sytem has been halted or paused.  During operations, we generate fake
triggers as needed to avoid saturation.

The SCA die measures 5 mm by 8 mm and was packaged in an 84 pin PLCC.
The yield was about 30\%.  Chips used in the forward TPC (FTPC)
readout were mounted in PQFP100 packages.

\subsection{FEE Card}

Fig. \ref{fFEEphoto} shows a photo of the FEE card.  The large chips
are the SAS and SCA's.  The card also includes a 90 MHz ADC clock and
a 2:1/4:1 multiplexer.  Signals are read out over a 50-pin ribbon
cable.  Most of the control signals are distributed over this cable,
along with regulated $\pm 5 $ VDC.  Control voltages for the SAS and
SCA are also distributed on this cable, but with on-board buffers for
the nodes with significant current draw.  The 90 MHz ADC digitization
clock is generated on the FEE card, to avoid high-frequency signal
distribution.

In addition to the pad signals, the 44-pin FEE connector also provides
an 8 bit geographical address.  These addresses are etched into the
pad plane, and uniquely identify each of the 181 FEE cards in a
sector.  These identifiers can be read out to the data acquisition
system through a geographical address event.  These geographical
events make it easy to identify cabling errors.

Since the TPC pad plane must be maintained to $\pm 0.7\deg$ C to keep
the wire gain constant and maintain the $dE/dx$ calibration, the FEE
cards needed to be water cooled.  The cards are cooled through a
mounting bracket which is glued and screwed to the back of the card.
This bracket mounts to a 3/8'' square aluminum cooling channel which
has water running through it.  This connection to the cooling channel
also serves as a ground.

Because the FTPC signals are slower, the FTPC system can be run at
lower currents.  Both the SAS input current and SAS buffer current are
significantly reduced.  This allows the FTPC FEE cards to be air
cooled.

\subsection{Noise}

Besides the SAS, noise comes from 3 sources: dielectric losses in the
TPC pad plane, charge injection noise in the SCA, and quantization
noise in the ADC.

The G-10 pad plane is a poor dielectric, with a loss angle
$\tan{(\delta)}\sim 0.014$.  The capacitance with a G-10 dielectric
acts as a noise source, generating noise equivalent to a resistor with
resistance given by the real part of the capacitance,
$R=\tan{(\delta)}/\omega C$.  For the STAR geometry and shaping, the
noise is about 300 electrons.

The SCA charge injection noise of 1 mV is equivalent to 420 electrons.
The SCA quantization noise is $1/\sqrt{12}$ ADC counts, or about 250
electrons.  All of the noise sources are independent, so the total
noise is their quadrature sum of 1075 electrons (for a relatively high
capacitance, 25 pF pad).

Fig. \ref{fFEEnoise} shows a histogram of the noise on the 5682
installed channels in one sector.  The mean is 1.3 counts, or 990
electrons, meeting the specifications; there is no significant
additional noise due to inadequate grounding, pickup, or other
external sources.

\subsection{Readout Boards}

Figure \ref{fRDO} shows a block diagram of the readout board.  Its
main function is as a multiplexer, reading data from the FEE cards and
sending it to the data acquisition system over a 1.2 Gbit/sec fiber
optic link.  Digitization on the FEE card proceeds synchronously with
data transmission, so there is little buffering on the readout board.

As Fig. \ref{frdophoto} shows, the FEE cards are divided into groups
of 4, 2 boards on each of 2 cables.  Each group is controlled by its
own field programmable gate array (FPGA), a Xilinx 4008.  Each group
also has its own positive and negative voltage regulators, and can be
turned on and off via the STAR control system.  Each FPGA is also
connected to a 40-bit wide 15 MHz tristate bus which transfers the
data to the serializer chip and laser diode.  The FPGAs contain state
machines which control the 4 boards, and buffers to synchronize the
FEE card readout with the data readout bus cycle.  The data are sent
to DAQ in an order that simplies the eventual on-line reconstruction.
The ordering matches the requirements of the data acquisition system ,
but is essentially arbitrary from the FEE standpoint. This required
ordering complicates the state machine.

The 40 bit wide bus feeds a 2:1 multiplexer/level shifter/latch which
itself feeds a HP G-link serializer which sends the data to a Finnisar
optical transmitter.  The G-link serializer runs at 60 MHz, so that
the data is sent to the DAQ system at 1.2 GHz; the link actually runs
at 1.44 GHz after Manchester encoding.  Because the DAQ system is less
than 100 meters from the detector, we use relatively inexpensive
50$\mu$m diameter multi-mode fiber for the transmission.

The readout board also has a 1 Mword 10-bit wide memory that can store
the data from a single event.  This memory is intended mainly for
diagnostic purposes, allowing the same event to be sent multiple
times.  It can also be written to and read out from the slow controls
link discussed below.

Data collection, digitization and readout is controlled by a trigger
bus, an 8 pair differential pECL bus that runs at 5 times the RHIC
crossing frequency (about 50 MHz); one cable drives an entire sector
of 6 readout boards.  Every 105 ns (each RHIC crossing), the trigger
transmits a 4-bit trigger action word, a 12-bit trigger token and a
4-bit DAQ action word.  The trigger action word can initiate readout,
abort the readout of an event in process (if it is rejected by a
higher level trigger), initiate calibration readout with the SAS
pulser, and write events to or read events from the event memory. The
token uniquely identifies each event, insuring that, during event
building, there is no mixing of pieces from different triggers.  The
DAQ word can be used to control DAQ functionality to allow for
different type of processing for different types of triggers.

The readout boards are monitored and controlled by a hardware controls
link\cite{controls}.  The system can monitor voltages, currents (via
the voltage drop in the power cables), and temperatures.  Thermistors
for temperature measurement are placed on the pad plane itself.  It
can also independently control power to each group of 4 FEE cards.
Finally, it can be used to read out the event memory.  Because this
readout takes 40 seconds/sector, is mainly for diagnostic purposes.

The link is based on the high-level data link control (HDLC)
protocol\cite{STARnote}.  The physical link is RS-485 running at 1
MBit/second.  On the readout board, the protocol is implemented in a
plug-in card holding a Motorola 68302 communications processor,
memory, and interface logic.  The off-detector end is a commercial VME
module.  Because of some limitations on the commercial modules, the
system includes 'looped back' timing, giving a system that behaves
closer to RS-422.  Each sector of 6 readout boards is fed by a single
HDLC cable.  The same hardware and protocol is used by several other
STAR subsystems.

The readout board is powered by $+8$ V and $-8$ V ferroresonant power
supplies.  A fully populated readout board consumes about 15 amps at
both $+8$ V and $-8$ V; each readout board is powered a by Kepco
Series C 300 watt power supply.  Ferroresonant supplies were chosen
for their reliability, moderate regulation and lack of switching
noise. They regulate output voltage to within 0.7 V (for 25\% to 100\%
load changes); linear regulators on the readout board provide better
regulation, but the initial regulation reduces the on-detector power
dissipation. The power cables are part of the system; $+8$ V, ground
and $-8$ V are each carried by two 12-gauge wires.  The ground lead
current is small because the positive and negative current consumption
are quite closely matched.  The voltages are monitored at the supply
and on the readout board; the voltage drop (typically about 1V) is
proportional to the current consumption. To avoid ground loops, the
cable grounds are not grounded at the power supplies.  The readout
boards are water cooled, using a similar setup as with the FEE cards.

There is no zero suppression on the readout boards.  The introduction
of relatively inexpensive, fast fiber optic links, together with
uncertainties about the algorithm for selecting non-zero sequences,
led STAR to implement zero suppression in the data acquisition system.

As Fig. \ref{frdophoto} shows, physically, the readout boards are
split into master and slave pieces.  The master contains all of the
unique parts of the circuitry: fiber driver, trigger interface, event
memory and slow controls interface, along with 5 FPGA FEE-card
controllers, enough to control 20 FEE cards.  The slave board contains
2, 3 or 4 FEE-card controllers, to control 8, 12 or 16 FEE cards.  This
multiple-sizing matches the readout boards to the number of nearby FEE
cards and available physical space.  On the inner sectors, the master
and slave boards are physically separated by a 20 cm cable.

The readout boards also have geographical addresses.  These are
set via programmed connectors which are physically tied to the
cooling structure.  

The tight space constraints on the FTPC led to some design changes in
the readout board.  Readout boards were mounted some distance from the
FEE cards, with cables up to 2 meters long, and connected with small
(0.025'') pitch cables.  To avoid digital crosstalk, the FEE cards
were read out at half the speed of the main TPC.  The optical link
speed was maintained with the use of double buffering.  A single board
layout was used, instead of the master/slave setup for the main TPC.

\subsection{Waveform Digitization Clock}

Although the readout boards operate quite independently, the SCA
signal acquisition must be synchronous across the entire detector to
simplify the event reconstruction.  The SCA acquisition frequency must
be adjustable to account for changing gas mixtures, TPC cathode
voltages, and possibly pressure.  At the same time, acquisition must
begin at a fixed phase with respect to the collision that causes the
trigger.  This is currently done by using the 10.5 MHz RHIC
beam-crossing clock as the SCA clock.  At this speed, the full drift
time corresponds to about 380 time-buckets.

In the future, a programmable clock may be used.  An analog prototype
based on a delay-locked loop was built, but it proved to be overly
temperature sensitive.  A programmable clock would allow the frequency
to be adjusted to make use of all 512 SCA time buckets.  The clock
will also be necessary if helium-ethane gas is ever used.

\section{Chip and Board Testing}

The system required 10,000 SAS and SCA chips, mounted on 5,000 FEE
cards.  With the yields, 15,000 SAS chips and 30,000 SCA chips had to
best tested.  Testing was thus a substantial fraction of the total
effort.

We developed a modular tester system, with similar circuitry used to
test both chips and the FEE cards.  Similar circuitry was used for
both chip and board debugging and for production testing.  We focus
here on the production testing.  Only a chip/board specific test card
was different for each component.  The testers were implemented on
PC's with National Instruments Lab-Windows.  The PC communicated with
the tester via a 12-channel MIO-16 analog board, with 12 inputs, each
digitized to 12 bits, and 2 DAC outputs.  Digital communication was
via a 96 channel DIO-96 digital interface board.

Much of the ciruitry for the testers was common to all 3 designs.  The
common circuitry included a programmable signal generator to produce
DC levels, pulses, and a sawtooth.  Control signals were generated by
FPGAs.

The testers were all double socketed.  The chip or board plugged into
a socket mounted on a small replaceable PC board, which plugged into a
socket on the tester.  This allowed for quick changes when the socket
wore out.

To keep track of the large number of chips and boards, each chip and
each FEE board was labelled with an individual bar code.  All testing
stations were equipped with bar code readers, allowing for automatic
identification and easy establishment of a test database. Test results
were stored in an EXCEL spreadsheet compatible format.

We decided to test the chips after they were packaged, eliminating the
difficult problem of probing the bare wafers.  The large chip count
required automated testers.  To speed the measurements, all 3 tests
sequences began with measurements of DC currents and voltages, to
quickly eliminate obviously bad components.

The SAS-specific board included a peak detector to measure the
amplitudes of output pulses, a peaking time detector to measure the
pulse rise time, a zero-crossing time detector to measure the fall time
(and check on the tail-correction circuitry), and a stair-case
generator to generate different sized input pulses.  The analog
outputs were fed to the MIO-16 board through a 16:1 multiplexer.
After the DC measurements, the tester checked for dead channels and
measured cross-talk by injecting pulses into each channel, and
observing the output in that channel and the two adjacent channels.
The RMS noise was measured, followed by the peaking and zero-crossing
times, and the signal width found.  Finally, the channel gains and
linearities were measured.  The complete test sequence for a good chip
took about 1 minute.

The SCA tester included a board to inject signals into the SCA at
adjustable times, to test different time buckets.  After DC tests, the
tester checked for dead channels and crosstalk by injecting pulses
into each channel at different times.  The tester looked for a
correctly timed signal in the pulsed channel and no signals
(i.e. crosstalk) in all of the other channels.  Next, the RMS noise of
the pedestal was calculated, and all channels and time-buckets scanned
for glitches (as from a single bad storage capacitor).  The gains and
linearity were then tested by injecting 10 DC levels into the SCA.
This test was used to classify the SCAs into groups based on their
gain.  To keep the SCA gains within a narrow (5\%) range, the groups
were used with different SCA ramp currents.  The ramp current was
adjusted by selecting one resistor on the FEE card.  Finally,
saw-tooth waveforms were injected into the SCA, and the results
displayed on the PC monitor, for a final visual inspection by the
operator.

The FEE card testing was quite similar to the SCA testing.  DC
currents and voltages were measured first.  These voltages and
currents were measured through a 10-pin test connector mounted on the
FEE card.  Channel functionality and cross-talk were checked.
Linearity/gain were measured for 10 different amplitude inputs; the
signals were checked at 3 different points on the waveform.  The RMS
noise was measured, and another glitch test performed.  Finally, the
pedestals were read out and displayed on the PC monitor, for operator
inspection.
 
\section{Conclusions}

We have designed, built and commissioned a 136,608 channel 6-12 MHz
10-bit waveform digitization system to read out the STAR time
projection chamber.  The electronics was installed on the TPC in
1998-99, and has been exercised extensively on cosmic rays and during
the summer, 2000 run.  The system is working as designed and
demonstrating excellent stability; over 99\% of the channels are
working within specifications.

We thank our STAR colleagues for their support.  The gas gain
measurements were done by Wayne Betts.  Sergei Panitkin provided
Fig. \ref{fFEEnoise}.  The SAS chip was designed by Eric Beuville, the
SCA primarily by Stuart Kleinfelder and the ADC by Oren Milgrome.
This work was supported in part by the Division of Nuclear Physics of
the Office of High Energy and Nuclear Physics of the U. S. Department
of Energy under Contract No. DE-AC-03-76SF00098.

\begin{table}
\begin{tabular}{lrr}
  & Inner Sector & Outer Sector \\
\hline
Pad Length & 11.5 mm & 19.5 mm \\
Pad Width  & 2.85 mm & 6.2 mm \\
Anode Wire-Pad spacing & 2 mm & 4 mm \\
Mean $N_{pri}$ & 33 & 79 \\
Wire Voltage & 1170 V& 1390 V\\
Wire Gain & 3770 &  1230 \\
Pad:Wire Coupling & 30\% & 34\% \\
Min. Ion Signal & 23,000 $e^-$   & 21,000 $e^-$ \\
Number of Pads & 1750 & 3942
\end{tabular}
\caption[]{Inner and outer sector parameters. The number of primary
electrons $N_{pri}$ is the number of electrons hitting the pad with
the largest signal, and depends on the geometry, drift, and pad size.}
\label{losses}
\end{table}

\begin{figure}
\caption{The readout chamber region of the STAR TPC. The gating grid
and ground plane wires are on a 1 mm pitch, while the anodes wires are
spaced every 4 mm.}
\label{fTPCRD}
\end{figure}

\begin{figure}
\caption{Overview of the on-detector STAR readout electronics. The
32-channel FEE cards plug into the TPC pad plane, while the readout
boards are mounted nearby.  The trigger, power and slow control cables
connect to the electronics platform, while the optical fiber goes to
the data acquisition system in a separate room.}
\label{foverview}
\end{figure}

\begin{figure}
\caption{A block diagram of the SAS chip.  Data acquisition is
initiated following a trigger when the reset switch is opened.  When
this happens, charge begins accumulating on $C_i$, and signals are
transmitted to the shaper.  The RC circuit $R_f$ and $C_f$ determine
the shaper time constant, while $R_p$ and $C_p$ determine the tail
correction.  Both $R_f$ and $R_p$ are electrically adjustable by
varying a control voltage.}
\label{fSAS}
\end{figure}

\begin{figure}
\caption{A block diagram of the SCA/ADC chip.  The switches connecting
the sample capacitors to the bus are controlled by a shift register.
The ADC ramp and 12 bit counter outputs are distributed to all 16 ADC
channels.}
\label{fSCA}
\end{figure}

\begin{figure}
\caption{A photograph of the FEE card. The right-hand connector plugs
into the pad plane, while the left connects to the readout board.  The
4 large chips are the SAS's (right) and SCA/ADCs (left). The input
protection resistors are also visible to the right.}
\label{fFEEphoto}
\end{figure}

\begin{figure}
\caption{A block diagram of the readout board. All of the unique
functionality is on the master, while the slave is merely an extension of the
40 bit bus, with additional control/FIFO/buffer units, to accommodate
additional FEE cards. The two halves may be plugged together, or connected 
by a short cable, to accommodate the outer and inner sector geometries
respectively.}
\label{fRDO}
\end{figure}

\begin{figure}
\caption{A photograph of a readout board. The master (left) contains
the trigger connector (upper left), laser diode (upper center) and
1-event memory (upper right).  Two small plug-in connectors for the
HDLC slow controls daughter card are visible between the trigger
section and the laser diode.  The 9 control/FIFO/buffer sections are
clearly visible on the bottom. The connector sticking up is the power
connection.}
\label{frdophoto}
\end{figure}

\begin{figure}
\caption{A histogram of the measured noise in one TPC sector. The
units are ADC counts; 1 ADC count $=$ 780 electrons.  These figures
match those measured on the bench.}
\label{fFEEnoise}
\end{figure}

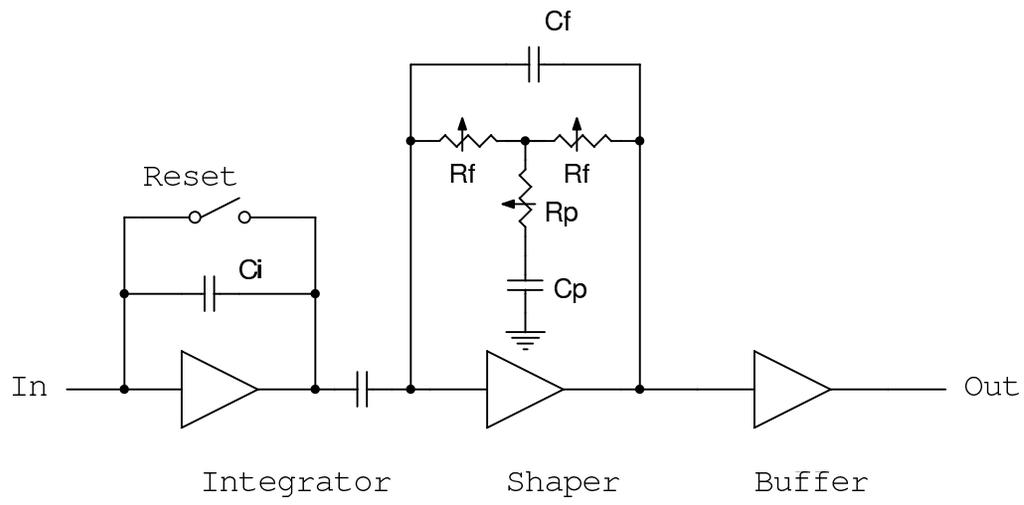
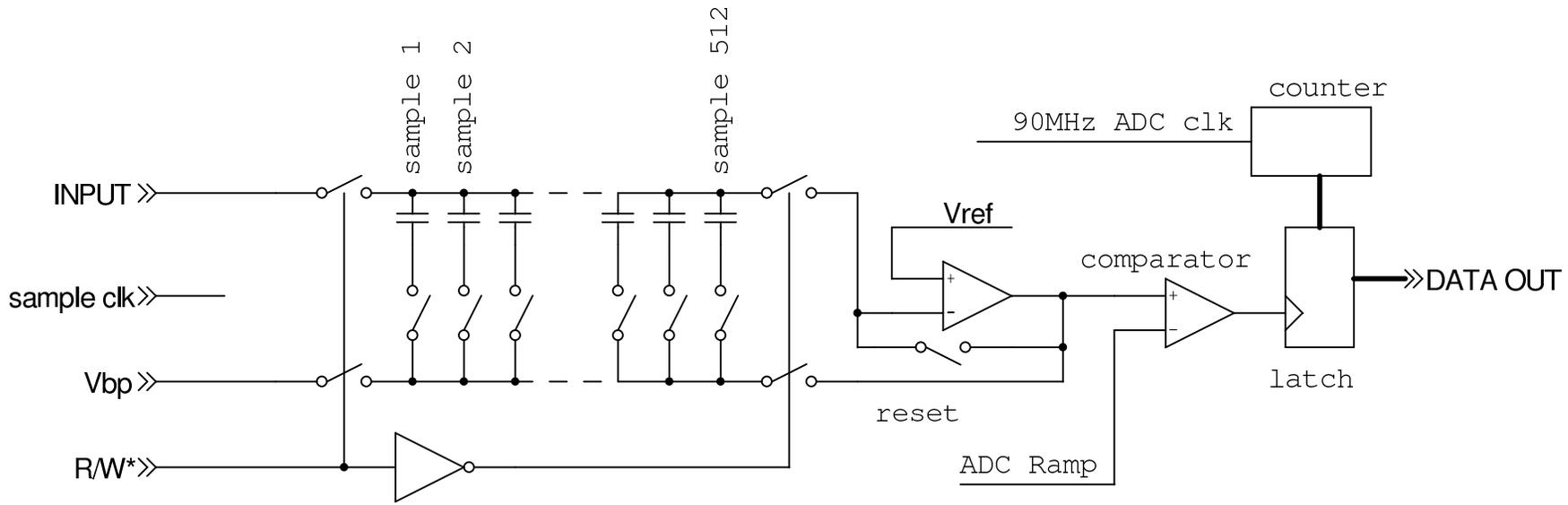
\end{document}